\documentstyle[12pt]{article}

\begin{document}

\topmargin 0pt
\oddsidemargin 5mm


\def\rv{{\bf r}}
\def\sv{{\bf s}}
\def\vv{{\bf v}}
\def\Ev{{\bf E}}
\def\Av{{\bf A}}
\def\Etild{\tilde{\bf E}}
\def\thetav{\vec\theta}
\def\gv{{\bf g}}
\def\kv{{\bf k}}
\def\qv{{\bf q}}
\def\Kv{{\bf K}}
\def\ex{0}\
\def\am{{\rm am}}
\def\vac{{\rm vac}}
\def\av{{\bf a}}
\def\bv{{\bf b}}
\def\nv{\hat{\bf n}}
\def\pol{{\bf\hat e}}
\def\sumtraj{\int_{\rm traj.}}
\def\xv{\hat{\bf x}}
\def\yv{\hat{\bf y}}
\def\zv{\hat{\bf z}}
\def\T{_{\rm T}}
\def\L{_{\rm L}}
\def\v{_{\rm v}}
\def\P{_{\rm P}}
\def\lex{l_{\rm e}}

\def\eg{{\it e.g.}}
\def\ie{{\it i.e.}}

\def\lambdabar{\lambda\raise0.4ex\hbox{\kern-0.5em\hbox{--}}\ }
\def\lambdaC{\lambda\raise0.5ex\hbox{\kern-0.5em\hbox{--}}_C\ }
\def\lesssim{<\kern-1.1em{\lower1.1ex\hbox{$\sim$}}\ } 
 \def\gtrsim{>\kern-1.1em{\lower1.1ex\hbox{$\sim$}}\ } 

\font\smallrm=cmr8 scaled 1000


\setcounter{page}{1}
\vspace{2cm}
\begin{center}

{\bf BUNCH COHERENCE IN PARAMETRIC X-RAY RADIATION}
\footnote{Contribution \`a : NATO Advanced Research Workshop
on Electron-Photon Interaction in Dense Media", (Nor Hamberd, Arm\'enie,
25-29 Juin 2001)}
\\
\vspace{1cm}
{\large  X.Artru $ ^{(a)} $ and K.A.Ispirian $ ^{(b)} $ }\\
\vspace{3mm}
{\em $ ^{(a)} $ Institut de Physique Nucl\'eaire de Lyon, IN2P3-CNRS and 
Universit\'e Claude Bernard, F-69622 Villeurbanne, France, \\
$ ^{(b)} $ Yerevan Physics Institute, Brothers 
Alikhanian 2, Yerevan, 375036, Armenia}\\

\end{center}

\vspace{1cm}
\centerline{\bf{Abstract}}

\medskip
\noindent
The amplitude of Parametric X-ray radiation emitted 
coherently by a bunch of electrons crossing a crystal (CPXR) is calculated.
When the bunch density is modulated with a 
longitudinal period close to the X-ray wavelenght, 
constructive many-electron interferences enhance the intensity by 
$N_b \times |F(\Kv)|^2$,
where $N_b$ is the number of electrons in the bunch,
$F(\Kv)$ the bunch form factor and $\Kv$ a specified wave vector.
CPXR can be used to test the microbunching in a X-ray free-electron laser.

\vspace{1cm}
\noindent
{\bf 1. Introduction}

\medskip
\noindent
In radiation source using bunches of relativistic electrons, 
strong enhancement by a many-electron coherence effect occurs 
if the bunch is shorter than the emitted wavelength 
or if it is modulated with a period close to this wavelength. 
In the first case, the radiation intensity is proportional 
to the square of the number $N_b$ of electrons in the bunch,
instead of being linear in $N_b$. 
Thus we have enhancement by a factor $N_b$.
In the second case, one may consider that the ``macro'' bunch is made of $M_b$ 
micro-bunches, each having $n_b$ electrons ($N_b=M_b \cdot n_b$).
If the microbunches are shorter than the wavelength and longitudinally
separated by $\lambda/v$ ($v$ is the beam velocity in units $c=1$),
we expect an enhancement factor $n_b$ for each microbunch,
times an enhancement factor $M_b$ coming  the interference between all the 
microbunches, thus an overall enhancement factor $N_b$ again.
In the general case, the amplitude per bunch is proportional
to the Fourier transform of the spatial charge distribution of the bunch.
Thus microbunching (MB) is a way of obtaining more radiation from
a given beam current. 
It is a basic ingredient of free-electron 
lasers (FEL) at optical wavelenght
and  "stimulated amplification of spontaneous emission" 
(SASE) in the X-ray domain (see [1,2] and refs therein).
Other potential applications are
coherent transition radiation (CTR) in the microwave [3-5]
and X-ray (CXTR) [6,7] regions, 
and possibly coherent radiation of particles channeled in crystalline and 
nanotube undulators [8-11]. 

Alternatively, coherent radiations of
various types (transition radiation, undulator radiation, etc)
can be used for the measurement of the beam longitudinal structure
with a spatial resolution of the order of $\lambda$ (see ref.[12]).
In the last years microwave CTR has been used for the 
measurement of the MB parameters [3-5]. 
At future SASE X-ray FELs [1,2], 
it would be interesting 
to have a direct and independent diagnostic of the MB process.
For this purpose, coherent X-ray transition radiation (CXTR) 
has been proposed [6,7].
In the present paper we discuss an alternative scheme 
based on coherent parametric X-ray radiation (CPXR).

\bigskip

\noindent
{\bf 2. Recall about PXR}

\medskip
\noindent
Parametric X-radiation (PXR) occurs when a charged particle propagates in a
crystal. This radiation is generated by Bragg diffraction of the
virtual photons composing the Coulomb field of the particle [13].
It is quasi-monochromatic and peaked in a small angular cone
about the specular direction (obtained from the electron direction
by symmetry with respect to the atomic planes). 
This direction is at large angle from the 
electron beam, thus eliminating the bremsstrahlung background.
PXR does not come alone, but is accompanied by
Diffracted Transition Radiation (DTR) [14-16]~: 
transition radiation
produced at the entrance surface of the crystal also undergoes
Bragg diffraction. PXR and DTR are have similar shapes and
usually interfere.

The spectral-angular photon distribution 
of PXR + DTR having momentum $\kv$ and polarization vector $\pol$ is
$$
I \equiv \omega \ {dN_\pol \over d^3\kv} 
= {\alpha \over 4 \pi^2} \ |A|^2
\,.\eqno(1)
$$
Using the reciprocity theorem, one can write the
amplitude $A$ as [15-17]
$$
A = \sumtraj d\rv\cdot\Ev_{\qv,\pol^*}(\rv) \ \ e^{i\omega t}
\,,\eqno(2)
$$
where $\Ev_{\qv,\pol^*}(\rv) \ e^{-i\omega t}$ describes a 
plane wave {\it coming from the detector} 
with momentum $\qv = - \kv$ and polarization $\pol^*$,
plus its scattered wave. Thus the calculation of the radiation 
amplitude reduces to an ordinary wave scattering problem.
In the kinematical approximation, 
the total scattering wave can be decomposed in 5 plane waves~:

- $|\qv_0,\pol>$ (using Dirac's ket notation)
incoming in vacuum from the detector,

- $|\qv,\pol>$, inside the crystal, 
obtained from the latter by refraction at the surface
of the crystal, 

- $|\qv',\pol'>$, inside the crystal, 
obtained from the latter by reflection about an atomic plane,

- $|\qv_\gv, \pol'>$, inside the crystal, 
defined by $\qv_\gv\equiv\qv+\gv$, 
where $\gv$ is a reciprocal lattice vector.

- $|\qv'_0,\pol'>$, in vacuum, obtained from $|\qv',\pol'>$ 
by refraction at the surface of the crystal.

The first two waves give negligible contributions to (2).
The last three momenta, $\qv'$, $\qv_\gv$ and $\qv'_0$
make small angles with $-\vv$ and give large contributions.
It is not obvious a priori which one of these momenta, if any,
is involved in the bunch form factor. This question will be the
main one tackled in this paper.

We consider the case of Laue geometry, for ultrarelativistic electrons
($\gamma \gg 1$). In the kinematical approximation the amplitude is
$$
A^{\rm (Laue)} = - \, i \ \vv\cdot\pol' 
\ {\delta_\gv  \over\ h' - h_\gv} \times B \times e^{i\Phi_l} \,,
\eqno(3)
$$ 
with
$$
B = \left( 1-e^{-i h_\gv T} \right) 
\left( {1\over h_\gv} - {1\over h_0} \right)
- {\rm idem}\{h_\gv \to h'\}
\,.\eqno(4)
$$
In the above formulas, $T$ is the travelling time of 
the electron in the crystal,
$$
h_0 \equiv \omega+\vv\cdot\qv'_0 
\,\simeq\, \omega\ (\gamma^{-2} + \theta^2) /2
\,,\eqno(5)
$$
$$
h' \equiv \omega+\vv\cdot\qv' 
\,\simeq\, \omega\ (\gamma^{-2} + \theta^2 - \chi_0) /2
\,,\eqno(6)
$$
$$
h_\gv \equiv \omega+\vv\cdot\qv_\gv
\simeq h_0 + g \, [\sin\Theta - g/(2\omega)] 
+ \omega \, (\nv\cdot\vv / \nv\cdot\hat\qv_0) \, \chi_0 /2
\,,\eqno(7)
$$
$\theta \sim \gamma^{-1}$ is the angle between 
$\kv$ and the direction specular to $\vv$,
$\Theta$ the angle between $\qv_0$ (not $\qv$) 
and the atomic planes
and $\nv$ the unit vector normal to the surface. 
$\chi_0 \equiv \varepsilon - 1 = -\,\omega\P^2 /\omega^2$ is the 
average dielectric susceptibility,
$\chi_\gv$ the Fourier coefficient of the local dielectric
susceptibility $\chi(\rv)$, and finally
$$
\delta_\gv \equiv {1\over2} \ (\pol\cdot\pol') \ \chi_{\gv} \ \omega 
\,.\eqno(8)
$$

Standard PXR consists in retaining only the $1/h_\gv$ term,
which gives a photon yield linear in $T$, due to the fact that $h_\gv$
can vanish. 
Integrating over $\omega$, one obtains the well-known formula [18],
$$
\left( {dN \over d\Omega} \right)_{\rm PXR} \simeq
T \ {\alpha\over4\pi} \ {\omega_{\rm B}\over\sin^2\Theta}
\ |\chi_\gv|^2
\ {\ \theta_\perp^2 + \cos^2 (2\Theta) \ \theta_\parallel^2
\ \over |\gamma^{-2} + \theta^2 - \chi_0|^2 }
\,,\eqno(9)
$$
in which $\theta_\parallel$ and $\theta_\perp$ are defined relative to the
($\gv,\vv$) plane and $\omega_{\rm B}=g/(2\sin\Theta)$ is the mean frequency.
One should be aware, however, that for thin or mosaic crystals the 
other terms can be equally important [15,16].

For the Bragg geometry, we have 
$
A^{\rm (Bragg)} = \exp(i\,h_\gv\,T) \ A^{\rm (Laue)}
$. 
In the dynamical theory for the Laue case, 
one has just to replace in (3) and (4) 
$h_g$ by $h_1$ and $h'$ by $h_2$
with
$$
h_{1,2} = {1\over2} \ \left\{
h' + h_\gv \mp \sqrt{
(h'-h_\gv)^2 + 4
(\nv\cdot\qv'/\nv\cdot\qv) \,\delta_\gv \, \delta_{-\gv} }
\ \right\}
\,.\eqno(10)
$$
For the Bragg geometry the modification is somewhat different [15,16].

\bigskip

\noindent
{\bf 3. The bunch coherence effect}

\medskip
\noindent
The phase factor $e^{i\Phi_l}$ is specific of the $l^{th}$ electron
of the bunch.
Calling $t_l$ and $\rv_l$ the entrance (resp. exit) time and position
for the Bragg (resp. Laue) geometry,
we have the three equivalent expressions
$$
\Phi_l = \omega t_l + \qv_\gv \cdot \rv_l
= \omega t_l + \qv' \cdot \rv_l
= \omega t_l + \qv'_0 \cdot \rv_l
\eqno(11)
$$
($\rv=0$ is taken on the entrance (resp. exit) surface).
Writing the electron trajectory as
$$
\rv(t) = \vv \ t + \sv_l
$$
where $\sv_l$ is the relative position of the electron inside the bunch,
we have 
$
t_l = - (\nv \cdot \sv_l) / (\nv \cdot \vv)
$
wherefrom
$$
\Phi_l = \Kv \cdot \sv_l 
\eqno(12)
$$
with the three equivalent expressions
$$
\Kv = \qv_\gv  + {h_\gv \over \nv \cdot \vv} \ \nv
= \qv'  + {h' \over \nv \cdot \vv}  \ \nv
= \qv'_0  + {h_0 \over \nv \cdot \vv} \ \nv
\,.\eqno(13)
$$
This momentum $\Kv$ is the one which enters in the bunch form factor.
Assuming that all the electrons of the bunch have the same 
velocity $\vv$ and neglecting possible two-body correlations
between them, the bunch coherent intensity is given by
$$
I_{bunch} = I_{one \ el.} \times
\left[N_b + N_b \, (N_b-1) \ |F(\Kv)|^2 \right]
\eqno(14)
$$
with
$$
F(\Kv) = \int\int\int d^3 \sv \ \rho(\sv)
\ e^{i \Kv \cdot \sv}
\,,\eqno(15)
$$
$\rho(\sv)$ being the electron probability distribution 
normalized to unity ($F(0) = 1$).

\bigskip
\noindent
{\bf 4. Approximations for $\Kv$}

\medskip
\noindent
In the pure PXR limit, $h_g=0$, then $\Kv = \qv_\gv$.
One can interpret $ - \qv_\gv$ as the momentum of a virtual photon
of the electron Coulomb field, which, 
after Bragg diffraction, becomes the real photon $-\qv$
going {\it toward} the detector.
The strongest amplification occurs when all the electrons emit this
virtual photon with the same phase factor $e^{i\qv_\gv \cdot \sv_l}$,
i.e., the electrons lies on the planes 
$\qv_\gv \cdot \sv_l = 2n\pi$ + constant. 
For a more general bunch modulation it is natural to get the 
form factor $F(\qv)$. 

For thin or strongly absorbing crystal, the PXR peak at Re$(h_g)=0$ 
becomes broad and less dominant.
At large enough $\gamma$ a DTR peak coming from the $1/h_0$ term 
becomes dominant, although $h_0$ cannot strictly vanish. 
Then one may neglect $h_0$ in (13) and take $\Kv\simeq\qv'_0$.

At very large $\gamma$, we have 
$h_0/\omega$, $h'/\omega$ and $h_\gv/\omega \ll 1$
and we may use equally well $\qv_\gv$, $\qv'$ or $\qv'_0$
as argument of the form factor.

Note that in backward coherent transition radiation
the argument of the bunch form factor is also
$\Kv = \qv'_0  + h_0 / (\nv \cdot \vv) \ \nv$,
where $\qv'_0$ is the momentum of a photon coming from the detector
after reflection by the surface.

\bigskip
\noindent
{\bf 5. Discussion}

\medskip
\noindent
We have shown that, in principle, bunch coherence can be obtained in
PXR (+ DTR) as well as, for instance, in transition radiation,
and have specified the relevant bunch form factor.
This phenomenon can be applied to the diagnostic of micro-bunching 
or to the enhancement of a PXR source. 
This is different from the (much more ambitious) goal of 
building a PXR Free Electron Lasers [19], 
which should operate with long pulses of extremely high densities
($10^8 - 10^9 \ A/cm^2$).

The advantages of PXR-DTR are (i) there is no need for a monochromator
(ii) photons are emitted at large angle from the beam.
These make the device more compact.
Two difficulties might be encountered~:

\noindent
- in MB diagnostics, the bunch may be accompanied by
X-rays coming from the undulator which makes the micro-bunching.
These X-rays will be Bragg-reflected by the PXR radiator
and can make a large background.
It may possible to eliminate these X-rays by a magnetic chicane
or using their linear polarization~: if the Bragg angle is 
45 degrees and the atomic planes perpendicular to 
the undulator plane, the reflection coefficient for the 
undulator X-rays vanishes. 

\noindent
- $\Kv$ makes the angle $\theta\sim\gamma^{-1}$ with the beam.
If the bunch is modulated only in the longitudinal coordinate
and has a transverse size $r_T$ larger than $\lambda / \theta$,
then its form factor is strongly reduced.
If one lowers $\theta$ too much, the $PXR+DTR$ intensity vanishes 
like $\theta^2$ 
(the same problem occurs with bunch-coherent transition radiation).
It should be however possible to avoid this damping
by making the spatial modulation of the bunch {\it oblique},
the planes of maximum density being orthogonal to $\Kv$ and not to the 
velocity.
The necessary tilt of these planes could be obtained
by deflecting the beam at the angle $-\theta$ using an upstream magnet.

\bigskip
\noindent
This work has been supported by INTAS contract 97-30392.

\bigskip
\noindent
{\bf 6. References}

\medskip

\medskip\noindent
 1. Conceptual Design of a 500 GeV   Linear Collider with Integrated X-
Ray Laser Facility, Eds. R.Brinkman et al, DESY 1997-048/ECFA 1997-
182,1997.

\medskip\noindent
 2. J.Arthur et al, LCLS.,Design Study Report, SLAC-R-521, UC-414,
1998.

\medskip\noindent
 3. J.Rosenzweig, G.Travish and A.Tremaine, Nucl. Instr. and Meth. A
365 
(1995) 255.

\medskip\noindent
 4. Y.Liu et al, Phys. Rev. Lett., 80 (1998), LL18.

\medskip\noindent
 5. A.Tremaine et al, Phys. Rev. Lett, 81 (1998) 5816.

\medskip\noindent
 6. E.G.Gazazian, K.Ispirian, R.K.Ispirian and M.I.Ivanian,
Pisma v Zh. Eksp. Teor. Fiz., 70, 664, 1999.

\medskip\noindent
 7. E.G.Gazazian, K.Ispirian, R.K.Ispirian and M.I.Ivanian, 
Nucl Instr and Meth. B173 (2001) 160 

\medskip\noindent
 8. A.V.Korol, A.V.Solovyov and W.Greiner, J.Phys. G.: Nucl. Part. 
Phys. 24, L45, 1998.

\medskip\noindent
 9.
R.O.Avakian, L.A.Gevorgian, K.A.Ispirian and R.K.Ispirian,
Pisma Zh. Esp.i Teor Fiz, 68, 437, 1998.

\medskip\noindent
 10.
A.V.Korol et al, Intern. J. Mod. Phys E, 8, 49,1999.

\medskip\noindent
 11.
R.O.Avakian, L.A.Gevorgian, K.A.Ispirian and R.K.Ispirian,
Nucl Instr and Meth. B173 (2001) 112. 

\medskip\noindent
 12.
Proc. 3rd European Workshop on Beam Diagnostics and Instrumentation
for Particle Accelerators, DIPAC 97, Frascati, 1997, LNF-97/048(IR), p.195-
251.

\medskip\noindent
 13.
G.M. Garibian and C. Yang, 
Zh. Eksp. Teor. Fiz. 61 (1971) 930
[Sov. Phys. JETP 34 (1972) 495]~;
V.G. Baryshevsky and I.D. Feranchuk,
J. Physique 44 (1983) 913~.

\medskip\noindent
 14.
A. Caticha, Phys. rev. A40 (1989) 4322~;
A.P. Potylitsin and V.A. Verzilov, Phys. Lett. A209 (1995) 380.

\medskip\noindent
 15.
P. Rullhusen, X. Artru and P. Dhez,
{\it Novel Radiation Sources Using Relativistic Electrons},
World Scientific, Singapore, 1998.

\medskip\noindent
 16.
X. Artru, P. Rullhusen,
Nucl. Inst. Meth. B145 (1998) 1~; {\it ibid} B173 (2001) 16.

\medskip\noindent
 17.
V.G. Baryshevsky, I.D. Feranchuk, A.O. Grubich and A.V. Ivashin,
Nucl. Inst. Meth. A249 (1986) 306.

\medskip\noindent
 18.
I.D. Feranchuk and A.V. Ivashin,
J. Physique 46 (1985) 1981~;
H. Nitta, Nucl. Inst. Meth. B115 (1996) 401.

\medskip\noindent
 19.
V.G.Barishevsky, K.C.Batrakov and I.Ya.Dubovskaya, 
J. Phys. D 24, 1250,1991~;
Nucl. Instr. and Meth., A375, 295, 1996.

\end{document}